\documentclass[useAMS]{mn2e}
\usepackage{epsfig}
\usepackage{rotating}

\def\lesssim{\,\lower2truept\hbox{${<\atop\hbox{\raise4truept\hbox{$\sim$}}}$}\,}
\def\gtrsim{\,\lower2truept\hbox{${>\atop\hbox{\raise4truept\hbox{$\sim$}}}$}\,}

\title[The PACO spectrally-selected sample]{The Planck-ATCA Co-eval Observations (PACO) project: the spectrally-selected sample}
\author[Bonaldi et al.]{
\parbox[t]{\textwidth}
{Anna Bonaldi$^{1}$\thanks{E-mail: anna.bonaldi@manchester.ac.uk}, Laura Bonavera$^2$, Marcella Massardi$^3$,  Gianfranco De Zotti$^{4,5}$}
\vspace*{8pt} \\
$^{1}$Jodrell Bank Centre for Astrophysics, School of Physics \& Astronomy, University of Manchester, Oxford Road, Manchester M13 9PL, U.K.\\
$^2$Instituto de F{\'\i}sica de Cantabria (CSIC-UC), Avda. los Castros s/n, 39005 Santander, Spain \\
$^3$INAF-Istituto di Radioastronomia, via Gobetti 101, I-40129 Bologna, Italy \\
$^{4}$INAF, Osservatorio Astronomico di Padova, Vicolo dell'Osservatorio 5, I-35122 Padova, Italy\\
$^{5}$SISSA, via Bonomea 265, I-34136 Trieste, Italy}
\begin{document}

\date{}

\pagerange{\pageref{firstpage}--\pageref{lastpage}} \pubyear{2012}

\maketitle

\label{firstpage}

\begin{abstract}
  The {\it Planck} Australia Telescope Compact Array ({\it Planck}-ATCA) Co-eval Observations (PACO) have provided multi-frequency (5--40 GHz) flux density measurements of complete samples of Australia Telescope 20 GHz (AT20G) radio sources at frequencies below and overlapping with {\it Planck} frequency bands, almost simultaneously with {\it Planck} observations. In this work we analyse the data in total intensity for the spectrally-selected (SS) PACO sample, a complete sample of 69 sources brighter than $S_{20\rm GHz}=200\,$mJy selected from the AT20G survey catalogue to be inverted or upturning between 5 and 20\,GHz. We study the spectral behaviour and variability of the sample. We use the variability between AT20G (2004--2007) and PACO (2009--2010) epochs to discriminate between candidate High Frequency Peakers (HFPs) and candidate blazars. The HFPs picked up by our selection criteria have spectral peaks $> 10\,$GHz in the observer frame and turn out to be rare ($<0.5\%$ of the $S_{20\rm GHz}\geq200\,$mJy sources), consistent with the short duration of this phase implied by the `youth' scenario. 
Most ($\simeq 89\%$) of blazar candidates have remarkably smooth spectra, well described by a double power-law, suggesting that the emission in the PACO frequency range is dominated by a single emitting region. Sources with peaked PACO spectra show a decrease of the peak frequency with time at a mean rate of $-3\pm 2\,\hbox{GHz}\,\hbox{yr}^{-1}$ on an average timescale of $\langle \tau \rangle = 2.1\pm 0.5\,$yr (median: $\tau_{\rm median}=1.3\,$yr).   The 5--20 GHz spectral indices show a systematic decrease from AT20G to PACO observations. At higher frequencies spectral indices steepen: the median $\alpha_{30}^{40}$ is steeper than the median $\alpha_{5}^{20}$ by $\delta\alpha = 0.6$. Taking further into account the Wide-field Infrared Survey Explorer (WISE) data we find that the Spectral Energy Distributions (SEDs), $\nu S(\nu)$, of most of our blazars peak at $\nu_{p}^{\rm SED}<10^5\,$GHz; the median peak wavelength is $\lambda_p^{\rm SED}\simeq 93\,\mu$m. Only 6 have $\nu_{p}^{\rm SED}>10^5\,$GHz.
\end{abstract}

\section{Introduction}\label{sect:intro}
The {\it Planck}--Australia Telescope Compact Array ({\it Planck}-ATCA) Co-eval Observations (PACO) project consists in observations with the Australia Telescope Compact Array (ATCA) of three {\it complete} samples drawn from the Australia Telescope 20 GHz (AT20G) catalogue (Murphy et al. 2010; Hancock et al. 2011) in three pairs of 2\,GHz-wide bands (5.5 and 9\,GHz, 18 and 24\,GHz, 33 and 39\,GHz). The selected samples are: the `bright PACO sample' including sources with $S_{20\rm GHz}\ge 500\,$mJy (Massardi et al. 2011b); the `faint PACO sample' (Bonavera et al. 2011) going down to $S_{20\rm GHz}= 200\,$mJy in the South Ecliptic Pole region; and the `spectrally-selected PACO sample', presented in this paper. The PACO observations were carried out in several epochs between July 2009 and August 2010. At least one observation was made within 10 days from the {\it Planck} satellite observations in any of the LFI channels (30, 44, 70\,GHz). Observations and data reduction procedures have been presented in Massardi et al. (2011b). The comparison between PACO data for all the samples and measurements in the {\it Planck} Early Release Compact Source catalogue (ERCSC; Planck Collaboration VII 2011) will be presented in Massardi et al. (in preparation).

A particularly interesting class of sources with spectra rising at frequencies above 5 GHz are the `High Frequency Peakers' (HFPs; Dallacasa et al. 2000), believed to be the most compact and youngest members of the Gigahertz Peaked Spectrum (GPS) source family (e.g., O'Dea\ 1998; Tinti \& De Zotti\ 2006). The first HFP selections (Dallacasa et al. 2000; Stanghellini, Dallacasa, \& Orienti 2009) had peak frequencies in the observer frame mostly in the range 5--10\,GHz. The AT20G survey offered the possibility of selecting complete samples of more extreme HFPs, with spectral peaks above 10 GHz (Hancock et al. 2010). The well established correlation between intrinsic peak frequency, $\nu_m$, and projected linear size (O'Dea 1998) indicates that $\nu_m \sim 10$--20\,GHz corresponds to scales $\sim 10\,$pc. Multi-epoch VLBA and VLBI observations of Compact Symmetric Objects (CSOs), as genuinely young HFP/GPS sources are generally found to be (Orienti et al. 2006; Orienti \& Dallacasa 2012), showed sub-relativistic expansion velocities, frequently of 0.1--$0.2c$ (Gugliucci et al. 2005; Polatidis 2009; An et al. 2012). Thus the AT20G HFPs are expected to have ages of a few hundred years and therefore to give us an insight on the initial stages of radio activity. The available data indicate that these early phases are characterised by high luminosity, fast evolution in size and peak frequency (Begelman 1996; Dallacasa 2003; Tinti \& De Zotti 2006), and must therefore be very rare. Only very large area, high frequency surveys, such as the AT20G that covers the entire Southern sky, can thus provide statistically significant samples or constraints.

A peaked spectral shape, however, does not guarantee that a source is truly young. Beamed radio sources (blazars) can also show peaked spectra when a flaring synchrotron component, self-absorbed up to high frequencies, dominates the emission spectrum. But the latter sources change their spectral shape keeping convex spectra for a relatively short time and may show a flat spectrum when re-observed after a few years, and anyway exhibit substantial variability (Tinti et al. 2005; Torniainen et al. 2007). In contrast, truly young sources maintain their convex spectra and show low or undetectable variability, apart from the spectral evolution associated to the source expansion (Orienti, Dallacasa, \& Stanghellini 2010). Thus multi-epoch, multi-frequency observations of peaked spectrum sources proved to be very effective in discriminating among the two populations. We use the variability properties between AT20G (2004--2007) and PACO (2009--2010) epochs to discriminate between young sources and blazars.

Our search for HFP candidates differs from the earlier study by Hancock et al. (2010), also based on the AT20G survey, in several respects. We start from a complete flux limited sample selected over the whole Southern sky, while the sample by Hancock et al. has no well defined flux density limit and is restricted to the region $18^h< \hbox{RA} < 01^h 30^m$. Also, their variability threshold for HFP candidates is mostly based on data at a single frequency (20 GHz) on relatively short timescales (1--3 yr), while we have multi-frequency data on somewhat longer timescales  (typically 2--4 yr, but up to 6 yr).  On the other hand, Hancock et al. obtained radio observations also at 95 GHz (while PACO observations span the range 5--40 GHz) as well as optical imaging and spectroscopy of a subset of their sources. As for blazars, the combination of AT20G and PACO data can constrain the duration of high frequency flares and the distribution of their peak frequencies. 

The structure of this paper is as follows. In \S\,\ref{sec:sample} we describe the data for the spectrally-selected PACO sample. In \S\,\ref{sec:blaz_vs_gps} and in \S\,\ref{sec:blazars} we discuss candidate HFPs and candidate blazars, respectively. Finally, in \S\,\ref{sec:conclusion} we summarise our main results.

\section{Sample characterisation}\label{sec:sample}
\subsection{Sample selection}\label{sec:selection}
The spectrally-selected (SS) PACO sample comprises all the 69 sources with $S_{20\rm{GHz}}>200$\,mJy and spectra classified in the AT20G catalogue (Massardi et al. 2011a) as inverted or upturning in the frequency range 4.8--20\,GHz, selected over the whole Southern sky. These sources have been re-observed for the PACO project between September 2009 and February 2010, with a scheduling process optimised to allow observations at all the frequencies almost simultaneous (i.e. within 10 days) with the {\it Planck} satellite. Fifteen SS sources are also part of the faint PACO sample ($S_{20\rm{GHz}}>200$\,mJy, Bonavera et al. 2011) and 21 of the bright PACO sample ($S_{20\rm{GHz}}>500$\,mJy, Massardi et al. 2011b), including 6 sources in common with the faint sample. 

The ATCA observations of each source over the frequency range 5--40 GHz have been carried out in the same day but, due to bad weather conditions, in one case (J173340-793555) a complete spectrum for the source has been obtained by joining the observations done in two subsequent days. The spectra of two other sources (J040019-225624 and J184827-735337) are not reliable as the sources turned out to be extended while the flux estimation relies on the point-like assumption. Data on these two sources are presented in the catalogue but not included in the analysis, so hereafter we consider a sample of 67 sources.

The classification of Massardi et al. (2011a) has been done by considering two spectral indices\footnote{We adopt the convention $S_\nu \propto \nu^\alpha$}: $\alpha_5^8$, between 4.86 and 8.56\,GHz, and $\alpha_8^{20}$, between 8.56 and 19.9\,GHz. At the time of the AT20G observations the sample was defined by having $\alpha_5^8>0$ and $\alpha_8^{20}>0$ (``inverted'') or $\alpha_5^8<0$ and $\alpha_8^{20}>0$ (``upturning''), after having excluded the sources with both  $|\alpha_5^8|<0.5$ and $|\alpha_8^{20}|<0.5$ (``flat'').

\begin{figure}
\begin{center}
\includegraphics[width=7cm,keepaspectratio,angle=90]{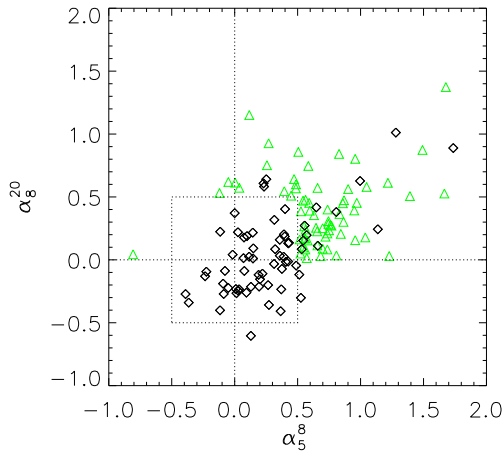}
\caption{Colour-colour diagram  ($\alpha_{8}^{20}$  versus $\alpha_5^{8}$) of the sources in the SS PACO sample. Green triangles: AT20G data on which the sample selection was made; black diamonds: PACO data. The dotted lines show the $\alpha_5^8=0$ and $\alpha_8^{20}=0$ axes and the region defined by $|\alpha_5^8|<0.5$ and $|\alpha_8^{20}|<0.5$ contains  sources classified as `flat-spectrum' by Massardi et al. (2011a).}
\label{fig:color-color_at20g}
\end{center}
\end{figure}

We have redone this classification using the PACO measurements over the 512 MHz-wide sub-bands centred at the frequencies nearest to the AT20G ones (4.732, 8.744 and 18.768\,GHz). The best quality PACO measurements, i.e. those referring to the epoch at which the highest accuracy was achieved (due to the combination of good weather, minimal radio interferences and instrumental malfunctions, etc.), were selected for this purpose. As illustrated by Fig.\,\ref{fig:color-color_at20g} a large fraction of the sample has significantly modified its spectral behaviour, moving to the ``flat-spectrum'' region of the diagram. This already shows that the majority of sources in the SS sample are blazars, caught in a bright stage by the AT20G survey, that decayed to a more quiescent phase at the time of PACO observations, i.e. on a timescale of a few to several years. We will come back to this in Sec.\,\ref{sec:blaz_vs_gps}, when we will discuss the selection of ``genuine'' HFP sources.

\subsection{Spectral behaviour}
\label{sec:pacoclass}
As a first step we have checked whether the spectra of our sources can be adequately described, over the full frequency range 5--40 GHz, by the double power-law
\begin{equation}\label{dpl}
S(\nu)=S_0/[(\nu/\nu_0)^{-a}+(\nu/\nu_0)^{-b}],
\end{equation}
used by Massardi et al. (2011a) and Bonavera et al. (2011) for the bright and faint PACO sample, respectively. $S_0$, $\nu_0$, $a$ and $b$ are free parameters. Such a smooth spectrum may not necessarily fit the data for blazars whose radio emission results from the superposition of peaked spectra of several parts of the jet racing towards the observer at highly relativistic speeds and self-absorbed up to different frequencies (e.g. De Zotti et al. 2010). The fit has been performed in logarithmic units by minimising the $\chi^2$  with a nonlinear optimisation technique based on an implementation of the Generalised Reduced Gradient optimisation method. Unless otherwise noted, the results reported below refer to the ``best quality'' PACO spectra (see \S\,\ref{sec:selection}).

\begin{figure*}
\begin{center}
\includegraphics[width=5cm,angle=90]{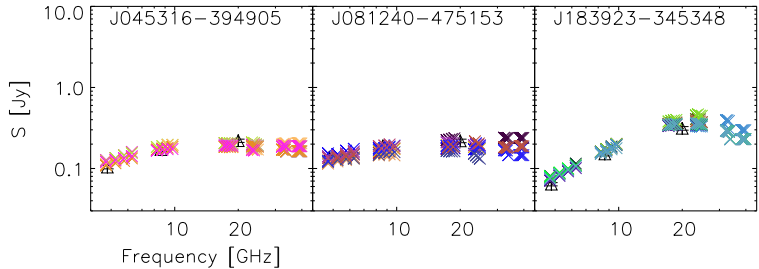}
\caption{PACO data (crosses) for the sources identified as HFP, with different colours corresponding to different epochs.  The dotted lines are the spectral fits the to the PACO data for the epoch with the most accurate data. Also shown for comparison are the AT20G data (triangles).}
\label{fig:gps}
\end{center}
\end{figure*}

We have considered as bad fits those with $\chi^2-\langle \chi^2\rangle>3\sigma_\chi$ where $\langle \chi^2\rangle$ and $\sigma_\chi$ are mean and standard deviation of the Gaussian fit of the reduced $\chi^2$ distribution over the whole PACO sample. Those are respectively  $\langle \chi^2\rangle=1.3$ and $\sigma_\chi=0.54$. We note that, in general, the $\chi^2$ values associated to the best fits are relatively high. This is due to the fact that, over the wide frequency range considered, the observed spectra show some small roughness, detected by our high accuracy measurements, so that a smooth function cannot provide a perfect fit. Still, the double power-law fits look acceptable for most sources (see Figs.~\ref{fig:gps}--\ref{fig:blaz_fl}).

For the SS sample, the double power-law model has been accepted for 57 out of 67 sources, i.e. in 85\% of the cases. For them we have computed low- (5--10\,GHz) and high-frequency (30--40\,GHz) spectral indices based on the fit and found that they fall in the following spectral categories:
 \begin{itemize}
 \item ``peaked'': $\alpha_{5}^{10}>0.3$, $\alpha_{30}^{40} <-0.3$; 20 sources (30\%)
 \item ``down-turning'': $\alpha_{30}^{40}\le \min(\alpha_{5}^{10}-0.35, -\alpha_{5}^{10})$ and $\alpha_{5}^{10}\le 0.3$; 26 sources (39\%)
 \item  ``self-absorbed'': $\alpha_{5}^{10}-0.35\ge \alpha_{30}^{40} \ge \max(-0.3, -\alpha_{5}^{10}$); 7 sources (10\%)
 \item ``flat'' with $|\alpha_{5}^{10}|$, $|\alpha_{30}^{40}| \leq 0.3$: 4 sources (6\%).
 \end{itemize}
No source switched from ``inverted'' or ``upturning'' according to the AT20G data to ``steep'' ($\alpha_{5}^{10}$, $\alpha_{30}^{40} <-0.3$) according to the PACO observations. Nor any source kept an ``inverted'' spectrum ($\alpha_{5}^{10}$, $\alpha_{30}^{40} >0.3$) up to the highest frequencies in the PACO range. A similar result was found for the bright and faint PACO samples. The peak frequency $\nu_p$ can be derived from the fitting parameters as $\nu_p=\nu_0(-b/a)^{(1/(b-a))}$. 

The remaining 10 (15\%) sources have complex spectra (see~\S\,\ref{sec:blazars}): besides a main peak, they also exhibit a sort of plateau, which can either be at $\nu$ lower (6 sources) or higher (4 sources) than the peak frequency. This behaviour could be indicative of multiple emission components.

\begin{table}
\begin{tabular}{ccc}
\hline
Frequency&PACO-AT20G&PACO-PACO \\
$[$GHz$]$&(2-4 yrs)&(6 months)\\
\hline
5&$13.1 \pm 1.9$&$5.5 \pm 1.5$\\
9&$15.0 \pm 2.2$&$4.9 \pm 1.7$\\
18&$20.3 \pm 2.5$&$5.7 \pm 1.7$\\
24&&$8.2 \pm 2.0$\\
33&&$4.8 \pm 3.0$\\
39&&$5.3 \pm 3.2$\\
\hline
\end{tabular}
\caption{Median variability indices [\%] of the SS sources and associated errors ($1.25\, \sigma/\sqrt{N-1}$, where $N$ and $\sigma$ are the number of the variability indices and the standard deviation of their distribution, respectively).}
\label{tab:variability}
\end{table}

\subsection{Variability}
Following Sadler et al. (2006) we define the variability index at the frequency $\nu$ as:
\begin{equation}
V^\nu_{\rm rms}=\frac{100}{\langle S^\nu \rangle}\sqrt{\frac{\sum [S^\nu_i-\langle S^\nu \rangle]^2-\sum(\sigma^{\nu}_i)^2}{n}},
\label{variability}
\end{equation}
where $S^\nu_i$ and $\sigma^\nu_i$ are the flux density and the associated error of a given source measured at the $i$-th epoch, $n$ is the number of epochs, and $\langle S^\nu \rangle$ is the mean of the available measurements. For 10--15\% of the sources (depending on frequency and time lag) $\sum [S^\nu_i-\langle S^\nu \rangle]^2<\sum(\sigma^{\nu}_i)^2$, so that the argument of the square root in eq.~(\ref{variability}) is negative. In these cases we estimated the $1\,\sigma$ upper limit to the variability index of the source as:
\begin{equation}
\sigma^\nu_{\rm var}=\frac{100}{\langle S^\nu \rangle}\sqrt{\frac{\sum(\sigma^{\nu}_i)^2}{n}}.
\label{variability_inf}
\end{equation}

The comparison between AT20G (2004-2008) and PACO (2009-2010) measurements gives information on the source variability at $\simeq 5$, 8, and 20 GHz ($V^5_{\rm rms}$, $V^8_{\rm rms}$, $V^{20}_{\rm rms}$) on typical timescales of $\sim 2$--4 years (but of up to 6 years in some cases), while the comparison of PACO measurements at different epochs informs on variability on a timescale of $\sim 6$ months, since they were coupled to the {\it Planck} scans which have a $\simeq 6$ month periodicity. The variability indices on the latter timescale have been computed, for all PACO frequencies, selecting for each source the best (smallest error bars) pair of observations spaced by 6 months within $\pm 20\%$.

The median PACO-AT20G and PACO-PACO variability indices and the associated errors are reported in Table\,\ref{tab:variability}. The PACO-AT20G variability of the SS sample turns out to be systematically higher than that found for the bright PACO sample ($11.8 \pm 2.1$, $11.7 \pm 2.2$ and $13.9 \pm 1.9$ for 5, 8 and 20\,GHz respectively; Massardi et al. 2011b), as expected since the SS sample contains a larger fraction of flaring objects. There is also a trend towards higher variability at higher frequencies, also found for the bright PACO sample (Massardi et al. 2011b). On the other hand the PACO--PACO variability amplitude is similar to that found for the bright sample on the same timescale (6 months). No trend with frequency shows up in this case. Of the three spectral subclasses containing at least 10 objects, the ``peaked'' one turns out to be the least variable.  On the 6 months timescale their median variability index is in the range 1.6--7.6, depending on frequency; the median variability index of ``down-turning'' sources ranges from 6.4 to 15.0 and that of ``complex spectrum'' sources ranges from 4.7 to 17.9. On the AT20G--PACO timescale (2--4 yr) the median variability indices are in the range 9--10, 13--32 and 15--40 for ``peaked'', ``down-turning'' and ``complex spectrum'' sources, respectively.

For 15 out of the 20 ``peaked'' sources we could determine the timescale for the variation of $\nu_p$ in the source frame, $\tau = \nu_p/(\Delta \nu_p/\Delta t_{\rm rest})$. Here $\nu_p$ is the peak frequency at the first epoch at which it could be determined, $\Delta \nu_p$ is its variation after a time $\Delta t_{\rm obs}$, in the observer frame, and $\Delta t_{\rm rest}=\Delta t_{\rm obs}/(1+z)$ is the corresponding time lag in the source frame. The redshift $z$ was available for 3 out of the 15 sources; for the remaining ones we used the median value over the candidate blazars of the sample having a redshift estimation ($z=1.33$). There is a trend towards a decrease of $\nu_p$ with time: $\Delta \nu_p/\Delta t_{\rm rest}$ is negative for 12 of the 15 sources; its mean value is $-3\pm 2\,\hbox{GHz}\,\hbox{yr}^{-1}$ and its median value is $-2.9\,\hbox{GHz}\,\hbox{yr}^{-1}$. This trend is expected if the size of the flaring region expands, and correspondingly its self-absorption frequency decreases, as it flows along the jet. The average variation timescale is $\langle \tau \rangle = 2.1\pm 0.5\,$yr and the median is $\tau_{\rm median}=1.3\,$yr. There is no detectable correlation between $\tau$ and $\nu_p$; this is not surprising since a correlation would anyway be hard to detect given the limited range of peak frequencies of our sources.

\section{Selection of HFP candidates}\label{sec:blaz_vs_gps}
\begin{table*}
\caption{HFP candidates. }
\label{tab:gps}
\begin{tabular}{lllllllllll}
\hline
AT20G name & RA [deg] & DEC [deg] & $\alpha_5^{10}$&$\alpha_{30}^{40}$&$\nu_p$\,[GHz]&z$^{(*)}$&$\nu_m$\,[GHz]&$l$\,[pc]\\
\hline
 J045316-394905            &73.3191&-39.8182&  0.61   & -0.16 & 15.66 &0.6&24.3&3.5\\
 J081240-475153            &123.171&-47.8647&  0.50   & -0.45 & 15.45 &0.7&26.4&3.1\\
 J183923-345348            &279.848&-34.8968&  1.39   & -1.05 & 21.62 &0.5&32.2&2.3\\
\hline
\multicolumn{9}{l}{$^{(*)}$ Estimated from B magnitude from superCOSMOS as in Burgess \& Hunstead (2006).}
\end{tabular}
\end{table*}
\begin{figure*}
\begin{center}
\includegraphics[width=12cm,angle=90]{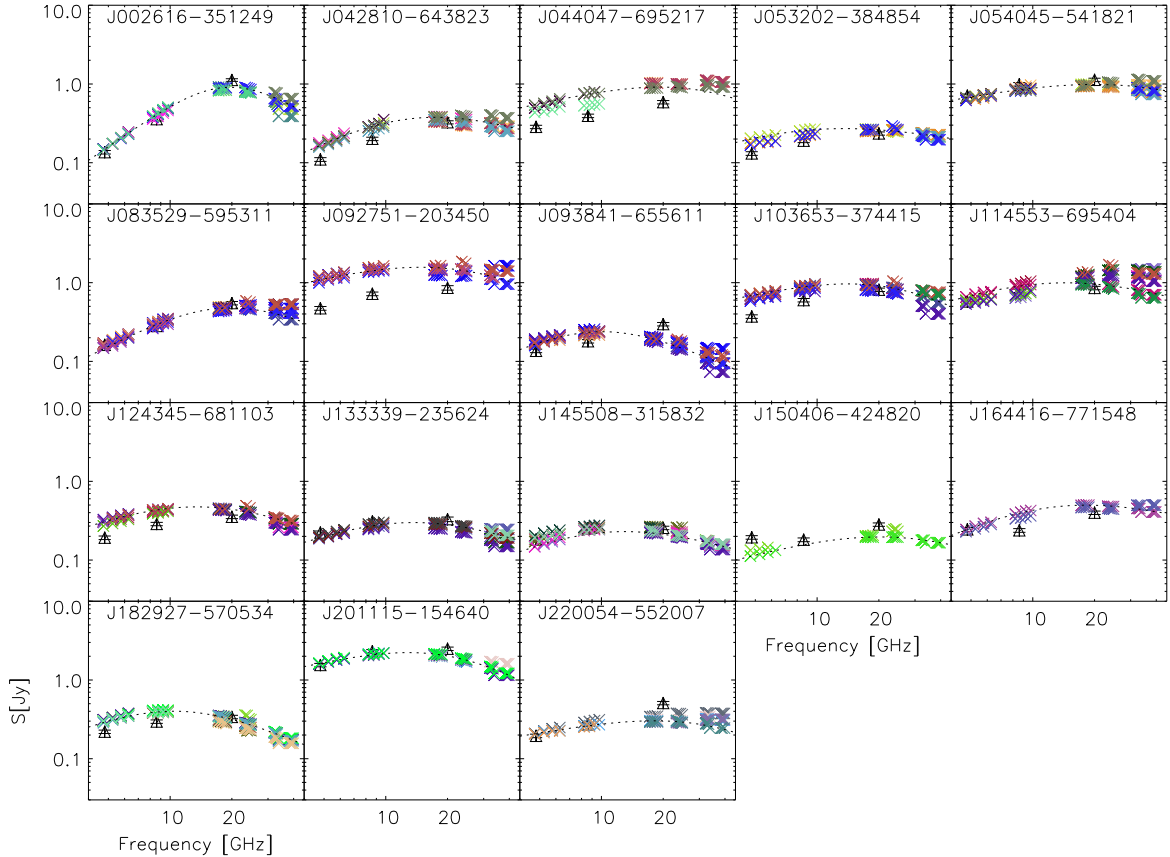}
\caption{PACO data (crosses) for the sources classified as spectrally peaked blazars with different colour corresponding to different epochs.  The dotted lines are the spectral fits to PACO data for the `best' epoch. Also shown for comparison are the AT20G data (triangles).}
\label{fig:blaz_pe}
\end{center}
\end{figure*}

\begin{figure*}
\begin{center}
\includegraphics[width=12cm,angle=90]{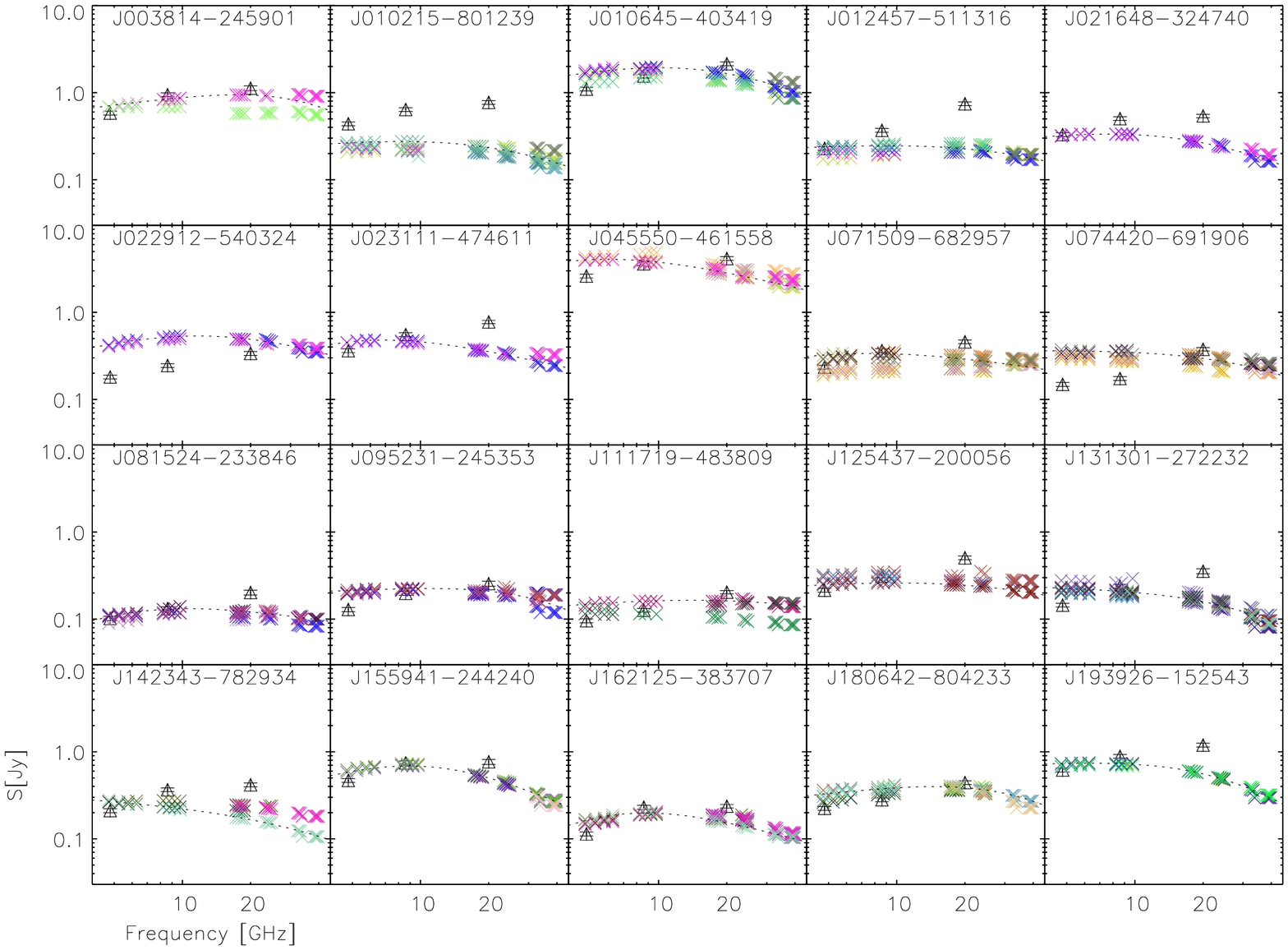}
\includegraphics[width=7.1cm,angle=90]{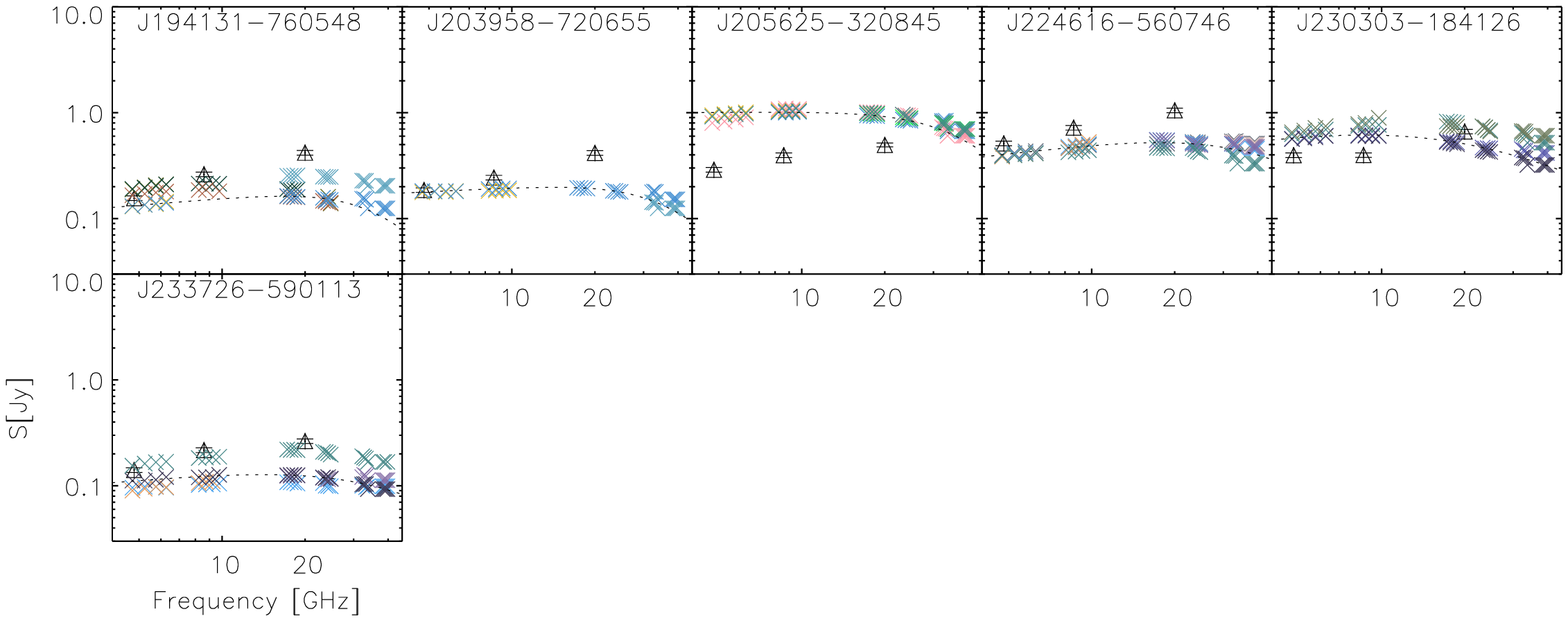}
\caption{Same as Fig.~\ref{fig:blaz_pe} for the sources classified as ``down-turning'' blazars. }\label{fig:blaz_ks}

\end{center}
\end{figure*}
\begin{figure*}
\begin{center}
\includegraphics[width=7.1cm,angle=90]{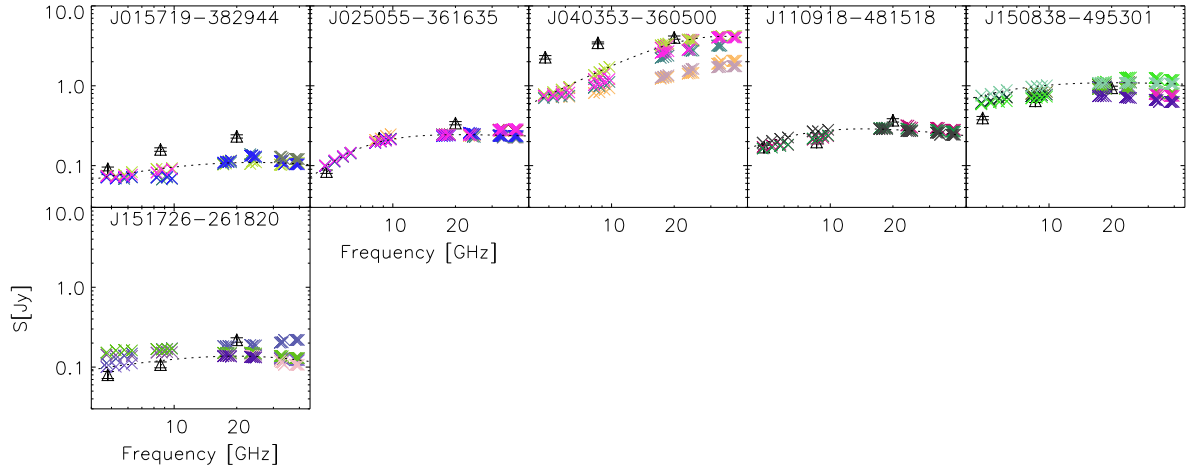}
\caption{Same as Fig.~\ref{fig:blaz_pe} for the sources classified as ``self-absorbed'' blazars.}\label{fig:blaz_ki}
\end{center}
\end{figure*}

\begin{figure*}
\begin{center}
\includegraphics[width=5.cm,angle=90]{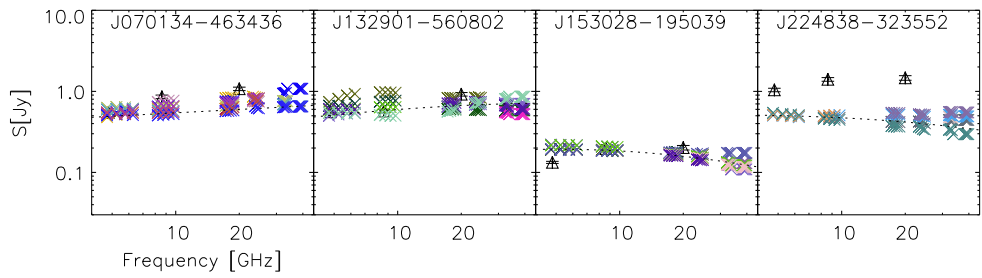}
\caption{Same as Fig.~\ref{fig:blaz_pe} for the sources classified as ``flat'' blazars.}\label{fig:blaz_fl}
\end{center}
\end{figure*}

\begin{figure*}
\begin{center}
\includegraphics[width=7.1cm,angle=90]{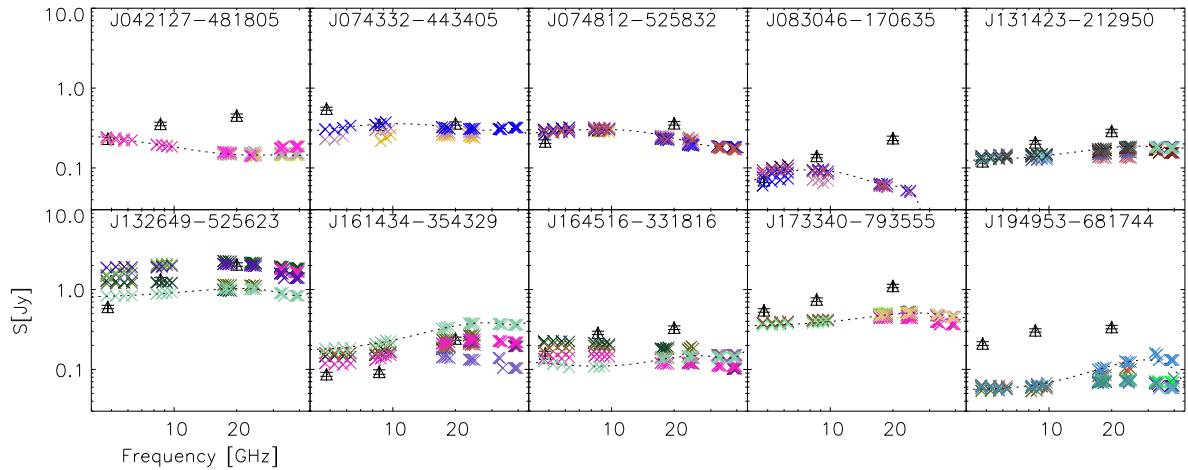}
\caption{Same as Fig.\ref{fig:blaz_pe} for the sources identified as blazars and classified as ``complex spectrum''. A polynomial fit is shown instead of the double power-law one (dotted line).}\label{fig:blaz_no}
\end{center}
\end{figure*}

\begin{table*} 
\caption{Blazar candidates in our sample. The blazar type is taken from Massaro et al. (2009) for the 24 sources in BZCAT. The spectral type is based on PACO data (see \S\,\protect\ref{sec:pacoclass}).}
\label{tab:blaz}
\begin{tabular}{ cccccccc } 
\hline
AT20G name & RA [deg] & DEC [deg] & blazar type&redshift& spectral type& $\nu_p$\,[GHz]& $\nu_m$\,[GHz]\\
\hline
J002616-351249 &6.56828&-35.2135& & &peaked &  19.66&\\
J003814-245901 &9.56133&-24.9839&QSO RLoud &0.498 &down-turning &  17.45&26.13\\
J010215-801239 &15.5632&-80.2111&QSO RLoud &1.169 &down-turning &  8.17&17.73\\
J010645-403419 &16.6880&-40.5722&QSO RLoud &0.584 &down-turning &  10.20&\\
J012457-511316 &21.2391&-51.2211&QSO RLoud &1.104 &down-turning &  8.97&18.87\\
J015719-382944 &29.3322&-38.4957&  & &self-absorbed &  24.32&\\
J021648-324740 &34.2008&-32.7946&QSO RLoud &1.331 &down-turning &  7.63&17.79\\
J022912-540324 &37.3033&-54.0567&  & &down-turning &  11.73&\\
J023111-474611 &37.7990&-47.7700&QSO RLoud &0.790 &down-turning &  6.88&12.32\\
J025055-361635 &42.7309&-36.2765&  &0.725 &self-absorbed &  21.01&36.24\\
J040353-360500 &60.9740&-36.0839&QSO RLoud &1.423 &self-absorbed & 34.5 ($51.0^\star$) & 123.6\\
J042127-481805 &65.3645&-48.3014&QSO RLoud &0.527 &complex &  &\\
J042810-643823 &67.0453&-64.6399&  & &peaked &17.40 &\\
J044047-695217 &70.1997&-69.8715&  & &peaked &19.24 &\\
J045550-461558 &73.9616&-46.2663&QSO RLoud &0.852 &down-turning &  5.97&11.06\\
J053202-384854 &83.0089&-38.8151&QSO RLoud &1.315 &peaked &13.54&31.36 \\
J054045-541821 &85.1910&-54.3061&QSO RLoud &1.190&peaked&19.93&43.64\\
J070134-463436 &105.394&-46.5769&QSO RLoud &0.822 &flat &  \\
J071509-682957 &108.790&-68.4997&  & &down-turning &  7.57&\\
J074332-443405 &115.887&-44.5681&  & &complex &  \\
J074420-691906 &116.084&-69.3185&  & &down-turning &  3.91&\\
J074812-525832 &117.053&-52.9756&QSO RLoud &1.802 &complex &  &\\
J081524-233846 &123.853&-23.6462&  & &down-turning &  10.69&\\
J083046-170635 &127.694&-17.1098&  & &complex & & \\
J083529-595311 &128.871&-59.8865&&&peaked&19.80&\\
J092751-203450 &141.966&-20.5809&QSO RLoud &0.348 &peaked &14.82&19.97 \\
J093841-655611 &144.674&-65.9366&  & &peaked &10.06 & \\
J095231-245353 &148.132&-24.8983&  & &down-turning &  10.82&\\
J103653-374415 &159.223&-37.7375&QSO RLoud &1.821 &peaked &14.15&39.91 \\
J110918-481518 &167.329&-48.2552&  & &self-absorbed &  14.47&\\
J111719-483809 &169.333&-48.6360&  & &down-turning &  14.25&\\
J114553-695404 &176.473&-69.9012&  & &peaked &14.53& \\
J124345-681103 &190.938&-68.1842&  & &peaked &13.44& \\
J125437-200056 &193.655&-20.0156&QSO RLoud &0.959 &down-turning &  9.99&19.58\\
J131301-272232 &198.258&-27.3756&  & &down-turning &&\\
J131423-212950 &198.599&-21.4974&  & &complex &  &\\
J132649-525623 &201.705&-52.9399&  & &complex &  &\\
J132901-560802 &202.255&-56.1341&  & &flat &  &\\
J133339-235624 &203.413&-23.9403&  &0.756&peaked&13.71&24.08\\
J142343-782934 &215.932&-78.4930&QSO RLoud & &down-turning && \\
J145508-315832 &223.787&-31.9757& & &peaked& 12.93&\\
J150406-424820 &226.026&-42.8056&  & &peaked &20.19& \\
J150838-495301 &227.162&-49.8840&  & &self-absorbed &  20.73&\\
J151726-261820 &229.361&-26.3058&  & &self-absorbed &  19.91&\\
J153028-195039 &232.620&-19.8443&  &0.851 &flat &  &\\
J155941-244240 &239.922&-24.7112&QSO RLoud &2.813 &down-turning & 8.88&33.85 \\
J161434-354329 &243.642&-35.7249&  & &complex &  &\\
J162125-383707 &245.355&-38.6188&  & &down-turning &8.76 &\\
J164416-771548 &251.067&-77.2635&  &0.043 &peaked &18.23&19.02 \\
J164516-331816 &251.320&-33.3047&  & &complex &  &\\
J173340-793555 &263.418&-79.5988&QSO RLoud &0.877 &complex &  &\\
J180642-804233 &271.677&-80.7093&  & &down-turning &  14.70&\\
J182927-570534 &277.365&-57.0929&  &0.840 &peaked &10.41 &19.16\\
J193926-152543 &294.861&-15.4286&QSO RLoud &1.657 &down-turning &  6.25&16.60\\
J194131-760548 &295.380&-76.0967&  &0.146 &down-turning &  16.80&19.25\\
J194953-681744 &297.472&-68.2957&  & &complex &  &\\
J201115-154640 &302.815&-15.7778&QSO RLoud  &1.180&peaked&12.89& 28.11\\
J203958-720655 &309.992&-72.1153&  &0.995 &down-turning & 14.29&28.51 \\
J205625-320845 &314.104&-32.1461&  & &down-turning &  7.20&\\
J220054-552007 &330.228&-55.3355&  & &peaked &18.11& \\
J224616-560746 &341.569&-56.1295&QSO RLoud &1.325 &down-turning &  17.57&40.85\\
J224838-323552 &342.161&-32.5978&QSO RLoud &2.268 &flat &  &\\
J230303-184126 &345.763&-18.6906&Uncertain type &0.129 &down-turning &  8.64&9.75\\
J233726-590113 &354.362&-59.0205&  & &down-turning &  14.20&\\
\hline
\multicolumn{8}{l}{$^\star$ Computed taking into account {\it Planck}/ERCSC data.}
\end{tabular}
\end{table*}
As mentioned in \S\,\ref{sect:intro}, variability studies are a powerful tool to distinguish GPS (and HFP) sources from flaring blazars. GPS sources should typically vary by no more than 10\% (O'Dea 1998) while blazars are in general much more variable. Six objects (out of the 67 included in the sample, i.e. $\simeq 9\%$) are found to have $V^\nu_{\rm rms}\le 10\%$. However, a low variability on the limited timescales of our observations is not a sufficient condition for sources not to be blazars, since blazars may stay in a quiescent state for several years. In fact, additional data taken from the NASA/IPAC Extragalactic Database (NED) and from the Massaro et al. (2009)  blazar catalogue do show evidence of strong variability for 3 out of the 6 sources, that therefore are classified as blazars. Thus only 3 HFP candidates remain (J045316-394905, J081240-475153, J183923-345348), i.e. 4.5\% of sources in the sample. They are all close to the 20 GHz flux density threshold. We list their relevant properties in Table~\ref{tab:gps} while their spectra are shown in  Fig.~\ref{fig:gps}. Their peak frequencies, $\nu_p$, in the observer's frame, are in the range 15--22 GHz.

There is a well established correlation between the rest frame peak frequencies, $\nu_m=\nu_p(1+z)$, and the projected linear size $l$ (O'Dea \& Baum 1997)
\begin{equation}\label{eq:Odea}
\log(\nu_m/{\rm GHz}) \simeq -0.21(\pm 0.05)-0.65(\pm 0.05) \log(l/{\rm kpc}), \label{eq:size}
\end{equation}
implying that intrinsic peak frequencies $> 10\,$GHz correspond to scales of a few tens of parsecs or less. As mentioned in \S\,\ref{sect:intro}, GPS sources are generally found to have expansion velocities of 0.1--$0.2c$ so that sizes of $\sim 10\,$pc correspond to ages of hundreds of years.

It must be noted that the selection criteria for GPS/HFP sources are somewhat fuzzy. For example, the sample of 21 GPS/HFP  candidates selected by Hancock et al. (2010) includes 5 sources with $S_{20\rm{GHz}}>200$\,mJy not included in our sample because they have both  $\alpha_5^8<0.5$ and $\alpha_8^{20}<0.5$, and were therefore classified as ``flat-spectrum'' by Massardi et al. (2011a). Three out of these 5 candidates were found to have higher variability than expected for genuine GPS/HFP galaxies, one was classified as a likely and one as a possible GPS/HFP galaxy. This suggests that our HFP sample may be incomplete. Since the area covered by the Hancock et al. sample is about 30\% of the AT20G area, the incompleteness may amount to a factor of $\sim 2$. On the other hand, the amount of variability for the Hancock et al. sample has been mostly determined on the basis of observations at  only two epochs with time lags of 1-3 years and only at 20 GHz. As pointed out by Hancock et al. themselves, well monitored AGNs show only small flux density variations on 1--2 year timescales, with larger outbursts occurring roughly once every 6 years. Hence more data are necessary to assess the GPS/HFP nature of the two additional candidates.

On the whole, we may conclude that the fraction of HFPs in our complete sample of inverted-spectrum sources with $S_{20\rm{GHz}}>200$\,mJy, albeit uncertain, is $< 10\%$.  Thus the high-frequency peaking population is largely dominated by relativistically beamed objects (blazars). A similar conclusion was reached by Bolton et al. (2006) for a sample of 9C sources brighter than 25 mJy at 15\,GHz with spectra peaking at $\nu_p > 5\,$GHz. Since SS sources are only 4.9\% of the 1360 AT20G sources brighter than that flux density limit, the fraction of HFP candidates among AT20G sources is $<0.5\%$. The rarety of genuine HFPs is indicative of a very short duration of the HFP phase, as expected in the framework of the `youth' scenario for these sources. For example, the model by Begelman (1996) implies $\nu_m \propto \tau^{-\delta\epsilon}$ (Tinti \& De Zotti 2006) where $\tau$ is the source age, $\delta\simeq 0.65$ is the slope of the $\nu_m$--size relation [eq.~(\ref{eq:Odea})] and $\epsilon=3/(5-n)$, $n$ being the slope of the radial density profile ($\rho_e \propto r^{-n}$) of the medium surrounding the expanding source. For the values of $n$ ($n=1.5$--1.9) favoured by Begelman (1996), the evolution timescale of $\nu_m$ is within a factor of two of the source age, i.e. of hundreds of years. According to the model by Snellen et al. (2000) $\nu_m \propto \tau^{-5/18}$ and the evolution timescale of $\nu_m$ is only moderately larger.

\begin{figure*}
\begin{center}
\includegraphics[width=6cm,keepaspectratio,angle=90]{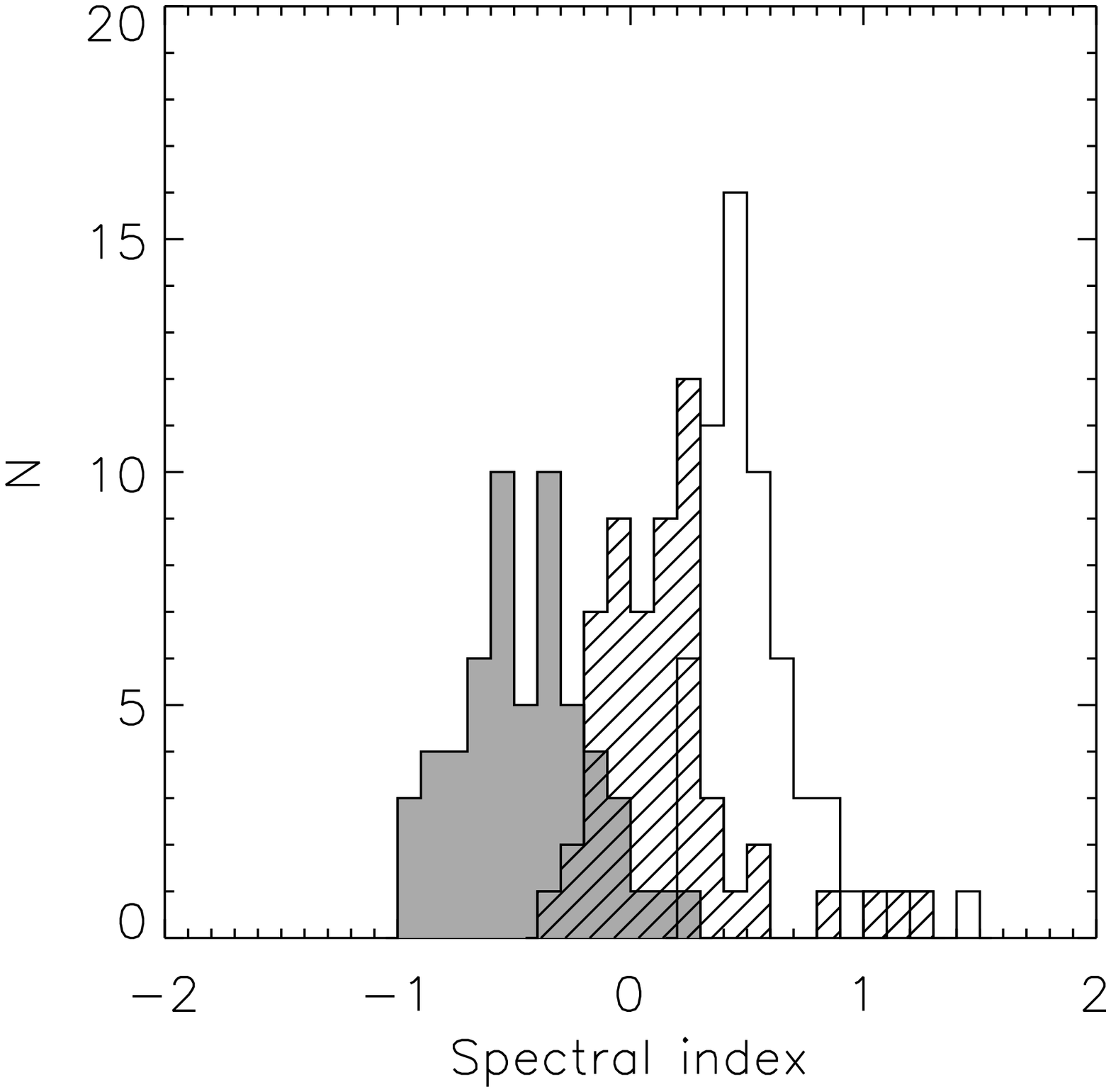}
\includegraphics[width=6cm,keepaspectratio,angle=90]{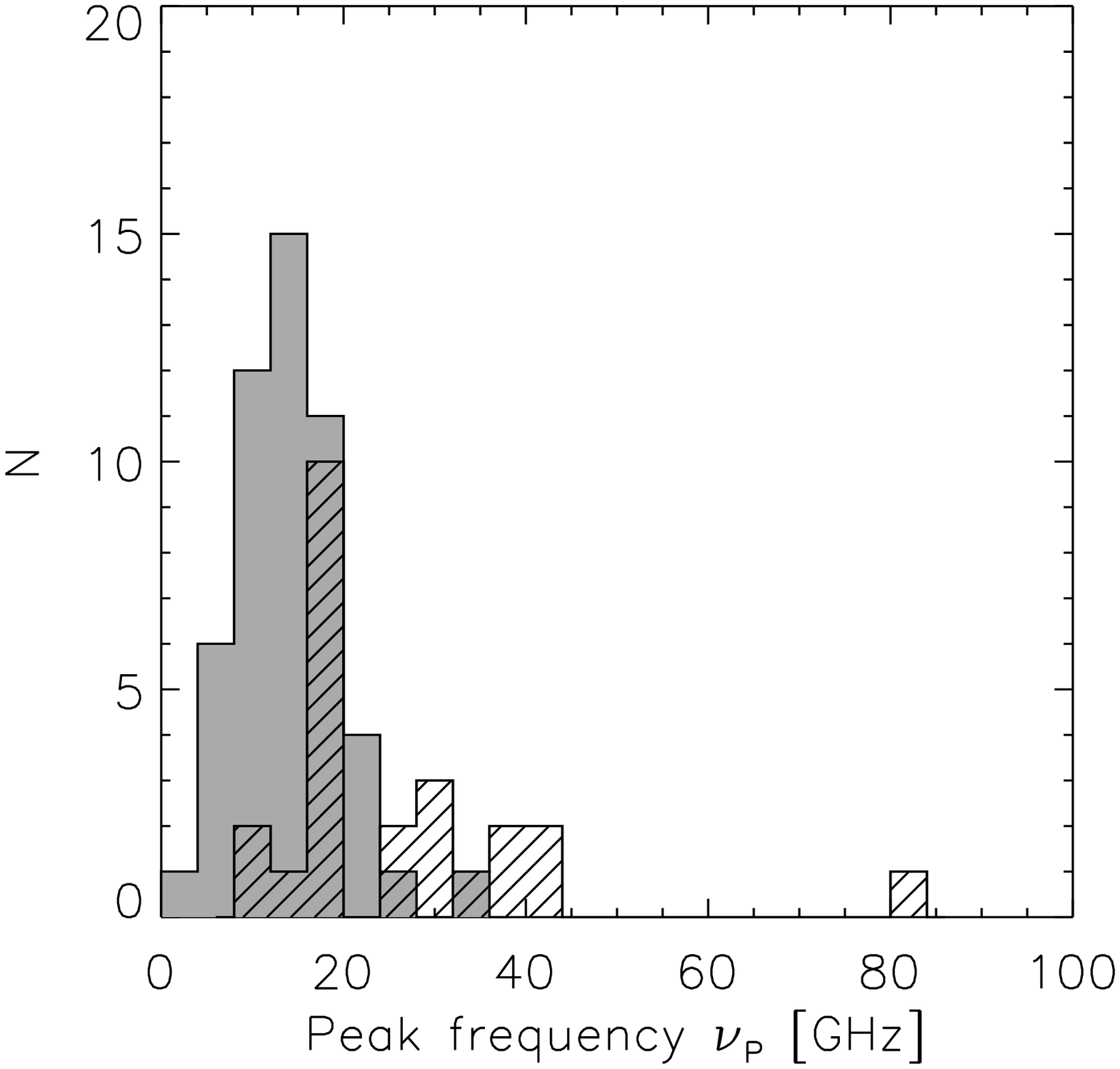}
\caption{Left: distribution of $\alpha_{5}^{20}$ (hatched histogram) and $\alpha_{20}^{40}$ (filled histogram) spectral indices for the sources with double power-law spectra, compared with the distribution of AT20G $\alpha_{5}^{20}$ for the same sample of sources (empty histogram). Right: distribution of peak frequencies for SS  sources with double power-law spectra (peaked, down-turning and self-absorbed) in the observer's frame (filled histogram) and in the rest frame for the subset of sources having a redshift measurement or estimate (hatched histogram).}
\label{fig:specinds}
\end{center}
\end{figure*}
\begin{figure}
\begin{center}
\includegraphics[width=6cm,keepaspectratio,angle=90]{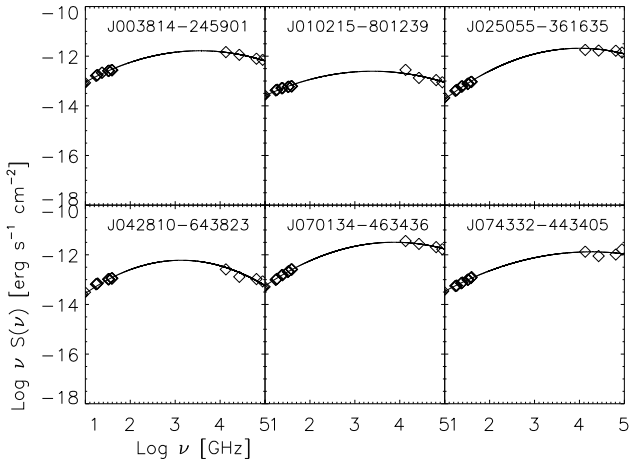}
\caption{Representative examples of SEDs including PACO and WISE data. Upper row, from left to right: two ``down-turning'' and a ``self-absorbed'' blazar. Lower row, from left to right: a ``peaked'', a ``flat'' and a ``complex spectrum'' blazar.}
\label{fig:sed}
\end{center}
\end{figure}
\begin{figure*}
\begin{center}
\includegraphics[width=6cm,keepaspectratio,angle=90]{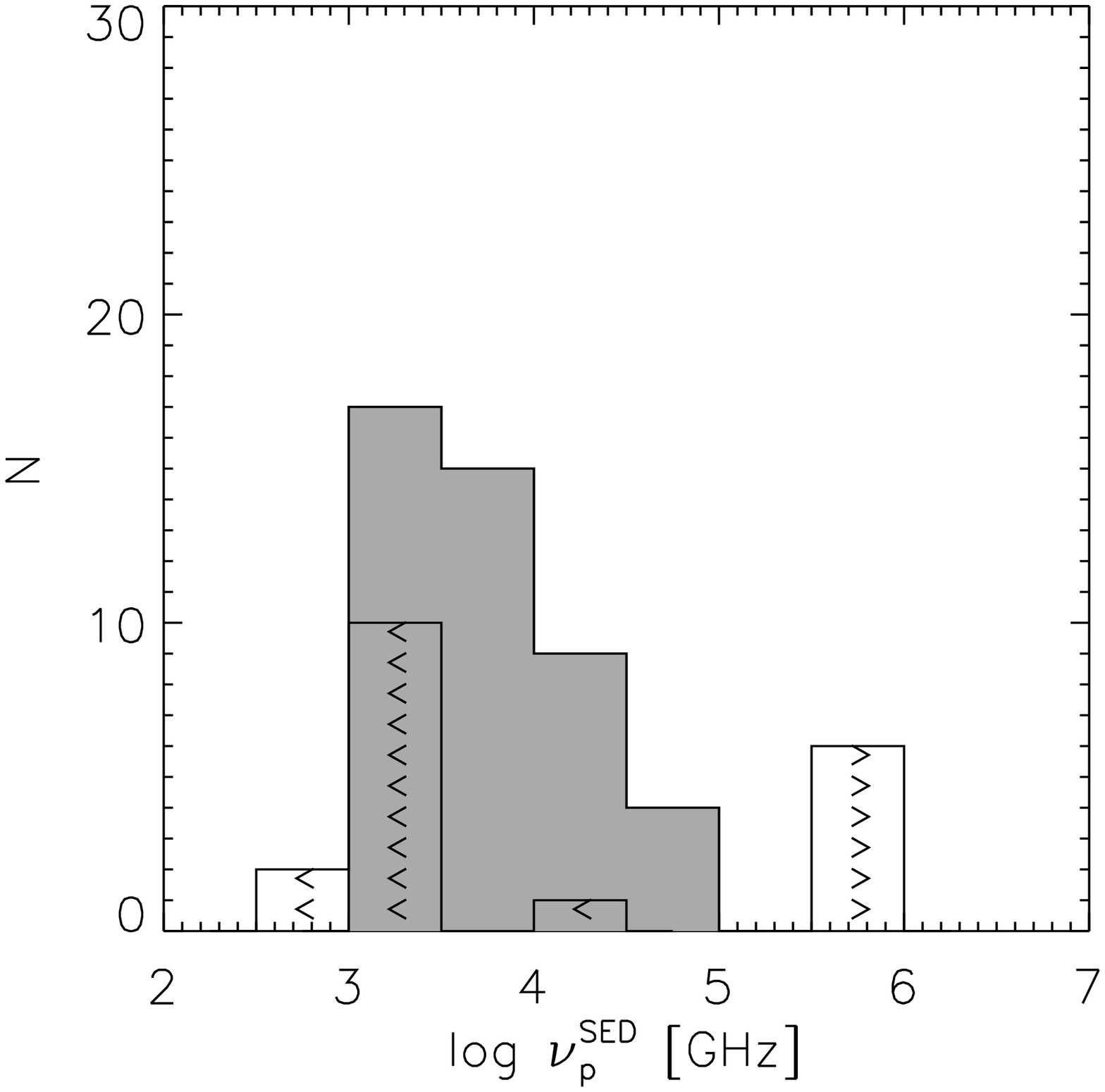}
\includegraphics[width=6cm,keepaspectratio,angle=90]{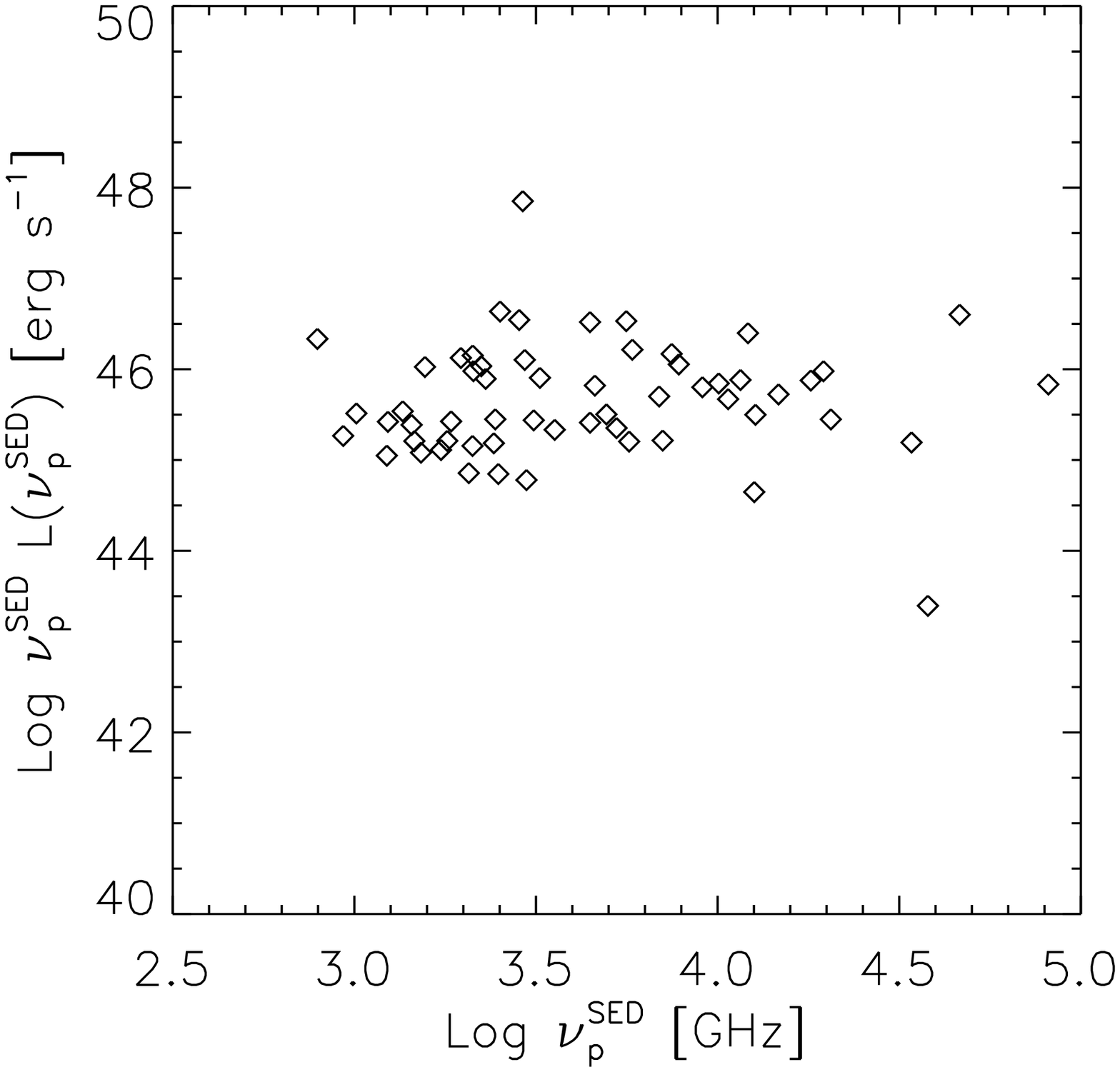}
\caption{Left-hand panel: distribution of the peak $\nu_{p,\rm SED}$ obtained by fitting a parabolic model to the PACO data combined with WISE data. The $>$ signs correspond to the 6 objects for which only a lower limit to $\nu_{p}^{\rm SED}$ could be obtained. The $<$ signs correspond to objects missing WISE photometry, for which we could obtain only upper limits to $\nu_{p}^{\rm SED}$. Right-hand panel: peak luminosity versus $\nu_{p}^{\rm SED}$.}
\label{fig:nupsed}
\end{center}
\end{figure*}

\section{Blazars}\label{sec:blazars}
All SS sources not classified as HFP candidates (64 sources, 95\%) are likely blazars. A positional cross-match with the Third Edition (April 2011) of the multi-frequency blazar catalogue (BZCAT) by Massaro et al. (2009) allowed us to confirm the blazar classification for 24 of them. 
The list of sources with coordinates and spectral properties is given in Table~\ref{tab:blaz}. The spectral classification and the value of $\nu_p$ are based on PACO data as described in Sec.~\ref{sec:pacoclass}. The blazar type is from the BZCAT where blazars are subdivided into flat-spectrum radio loud QSOs (QSO RLoud), BL Lacs and Uncertain type. With only one exception (Uncertain type) all our blazars included in BZCAT (which are however a minority, $\simeq 40\%$ of the total number of our blazars) are classified as QSO RLoud.

The spectra based on PACO observations are shown in Figs.~\ref{fig:blaz_pe}--\ref{fig:blaz_no} (one figure per spectral type) for all epochs, together with AT20G (triangles) data points. 
Even though the sources were selected for having inverted or upturning AT20G spectra, the PACO observations show, for most of them, flattish spectra in the 5--20 GHz range. The median PACO spectral index $\alpha_{5}^{20}$ is 0.14, to be compared with the median AT20G value of 0.48 for the same sources (see the left panel of Fig.~\ref{fig:specinds}). This indicates that these sources were caught by AT20G observations during a bright emission phase of a dense region probably in the inner part of the jet, and that the emission decayed by the time of PACO follow-up. The spectra show a significant steepening at higher frequencies (see again the left panel of  Fig.~\ref{fig:specinds}): the difference between the median $\alpha_{5}^{20}$ and the median $\alpha_{20}^{40}$ is $0.6$. Only one source, J040353-360500, has $\alpha_{20}^{40}>0$ ($\alpha_{20}^{40}=0.26$); its spectrum however flattens towards the top of the frequency range ($\alpha_{30}^{40}=-0.006$).


The distribution of peak frequencies of SS sources with double power-law spectra is displayed in the right-hand panel of Fig.~\ref{fig:specinds}. The mean peak frequency of blazars, in the observer's frame, is 14\,GHz with a standard deviation of 6\,GHz. The candidate HFPs lie in the low-frequency part of the distribution. The median rest-frame peak frequency of the 24 blazars with redshift information is 20\,GHz.

Interestingly, with only one exception, the `best epoch' spectra of sources for which a double power-law [eq.~(\ref{dpl})]  fit could be obtained have $\nu_p<40$\,GHz.  The exception is  J040353-360500 (see Fig.~\ref{fig:blaz_ki}) for which \emph{Planck} ERCSC data are available; a double power-law fit including them yields $\nu_p\simeq 51$\,GHz.

Since all SS sources have high-frequency spectral indices $\alpha_{30}^{40}>-1$ their SEDs $\nu S(\nu)$ are rising in the frequency range covered by PACO observations. To determine the SED peak frequency, $\nu_{p}^{\rm SED}$, we constrained the descending part of the spectrum with WISE (Wright et al. 2010) data, at 3.4, 4.6, 12 and 22\,$\mu$m ($\sim$88000--13600\,GHz). Of the 64 blazars, 51 (79\%) have a counterpart in WISE; for the remaining 13 (21\%) we used the upper limits given by Wright et al. (2010): 0.08, 0.11, 1 and 6\,mJy at 3.4, 4.6, 12 and 22\,$\mu$m respectively. We fitted the PACO plus WISE data with a parabola in the $\log \nu$--$\log \nu S(\nu)$ plane, parametrised by the $0^{\rm th}$, $1^{\rm st}$ and $2^{\rm nd}$ order coefficients $c_0$, $c_1$ and $c_2$, and got $\nu_{p}^{\rm SED}=-c_1/(2c_2)$. The accuracy of this estimation is limited by the gap between the PACO and the WISE frequency ranges and by variability, as PACO and WISE data are not simultaneous. In two cases the PACO plus WISE data turned out to be incompatible with a parabolic fit, but for the other objects the fits looks reasonably good. Some representative examples are shown in Fig.~\ref{fig:sed}.

The distribution of $\nu_{p}^{\rm SED}$ is shown in the left-hand panel of Fig.~\ref{fig:nupsed}. Most of them (56) have $\nu_{p}^{\rm SED}< 10^5\,$GHz and are therefore classified as ``Low Synchrotron Peak'' (LSP) according to Abdo et al. (2010). Their median peak frequency ($\log(\nu_p/\hbox{GHz})=3.51$) corresponds to a peak wavelength $\lambda_p^{\rm SED} \simeq 93\,\mu$m; there is no significant trend of the peak frequency $\log(\nu_p/\hbox{GHz})$ with the PACO spectral class. 

For the remaining 6 sources (J042127-481805, J093841-655611, J182927-570534, J194131-760548, J203958-720655, J230303-184126) the fit yields values of $\nu_{p}^{\rm SED}$ larger than $10^5\,$GHz, i.e. above the range covered by the data. Thus, only a lower limit ($\nu_{p}^{\rm SED}\gtrsim 10^5\,$GHz) could be obtained.  The Two Micron All Sky Survey (2MASS; Skrutskie et al. 2006) photometry, available for all these sources, confirms that their SEDs keep flat up to at least the J-band ($2.4\times 10^5\,$GHz). B-band photometry, given in the NED for 3 of them (J042127-481805, J093841-655611, J230303-184126), is consistent with SEDs being still flat. Thus, these are likely ``High Synchrotron Peak'' (HSP) sources ($\nu_{p}^{\rm SED}> 10^6\,$GHz). The sharp drop of the $\nu_{p}^{\rm SED}$ distribution below $\nu_{p}^{\rm SED}= 10^3\,$GHz is a selection effect: sources with lower values of $\nu_{p}^{\rm SED}$ are too faint to be detected by WISE and therefore we could only obtain upper limits ($\gtrsim 10^3\,$GHz) to their peak frequencies.

The addition of WISE data gives us a substantially different perspective on source spectra. Three of the 6 likely HSPs were classified as ``down-turning'' on the basis of PACO data, 2 other likely HSPs were classified as ``peaked'', and the remaining one was classified as ``complex spectrum''. This illustrates once more the complexity and the time-dependence of blazar spectra.

The right-hand panel of Fig.~\ref{fig:nupsed} shows $\nu_{p}^{\rm SED}$ versus the peak luminosity [$\nu_{p}^{\rm SED}\,L(\nu_{p}^{\rm SED})$]. According to the blazar sequence model (Fossati et al. 1998; Ghisellini et al. 1998), the synchrotron peak frequency is anti-correlated with the radio luminosity. Our data do not show any statistically significant correlation: the Spearman correlation coefficient is $\rho=-0.05$, implying that the null hypothesis (no correlation) is rejected only at the 0.4\,$\sigma$ level. 

\section{Conclusions} \label{sec:conclusion}
We have presented an analysis of the ``spectrally-selected'' (SS) PACO sample comprising all the AT20G sources with $S_{20\rm{GHz}}>200$ mJy and spectra classified by Massardi et al. (2011a) as inverted or upturning in the frequency range 5--20\,GHz. The sample consists of 69 sources, two of which were however rejected because they turned out to be extended and thus to have unreliable flux density estimates. The PACO observations were carried out in three pairs of 2\,GHz-wide bands (5.5 and 9\,GHz, 18 and 24\,GHz, 33 and 39\,GHz) in several epochs between July 2009 and August 2010. At least one observation was made within 10 days from the {\it Planck} satellite measurements.

Most (85\%) of the source spectra are remarkably smooth and well described by a double power-law. Twenty have a peaked spectrum. Most (12 out of 15) of those for which PACO observations allowed an accurate determination of the spectrum at different epochs show a decrease of the peak frequency with time. The mean decrease rate for the 15 peaked sources is $-3\pm 2\,\hbox{GHz}\,\hbox{yr}^{-1}$. The average variation timescale is $\langle \tau \rangle = 2.1\pm 0.5\,$yr and the median is $\tau_{\rm median}=1.3\,$yr.

Only three of our sources are found to be good candidate High Frequency Peakers, with peak frequencies, in the observer's frame, $\nu_p>10\,$GHz. Their 20\,GHz flux densities are close to the sample threshold and their peak frequencies lie in the lower part of the distribution for the full sample. Even allowing for a generous correction for the possible incompleteness of our selection, these extreme ($\nu_p> 10\,$GHz) HFPs are only a small fraction ($< 0.5\%$) of sources with $S_{20\rm{GHz}}>200\,$mJy, consistent with the very short duration of the HFP phase expected in the framework of the `youth' model.

All the other sources in our sample are likely blazars, and 24 of them, all but one classified as flat-spectrum radio QSOs, are listed in the last edition of the blazar catalogue by Massaro et al. (2009; BZCAT). We find a clear decrease of 5--20 GHz spectral indices from AT20G to PACO observations, implying that these sources were generally caught by the AT20G survey during a bright phase at 20 GHz. At higher frequencies spectral indices steepen: the median $\alpha_{20}^{40}$ is steeper than the median $\alpha_{5}^{20}$ by $\delta\alpha = 0.6$. All but one of the sources for which a peak of the flux density distribution could be defined have $\nu_p<40\,$.  This is at odds with the results by Hancock et al. (2010) who found that 12 of their 21 sources have $\nu_p > 40\,$GHz, including 4 sources peaking above 95 GHz, but is consistent with those by Massardi et al. (2011) who found that only a few of the 189 sources in their complete sample with  $S_{20\rm{GHz}}>500\,$mJy have $\alpha_{30}^{40}> 0$.

On the other hand, all SS sources have high-frequency spectral indices $\alpha_{30}^{40}>-1$ so that their SEDs $\nu S(\nu)$ peak above the frequency range covered by PACO observations. Taking into account the WISE photometry we find that the vast majority of them have $\nu_{p}^{\rm SED}<10^5\,$GHz and are therefore classified as LSPs, according to Abdo et al. (2010). Their median peak wavelength is $\lambda_p^{\rm SED} \simeq 93\,\mu$m. Only 6 of our blazars are likely HSPs.

\section*{Acknowledgements}
MM and GDZ acknowledge financial support from ASI/INAF Agreement I/072/09/0 for the {\it Planck} LFI activity of Phase E2. This publication makes use of data products from the Two Micron All Sky Survey, which is a joint project of the University of Massachusetts and the Infrared Processing and Analysis Center/California Institute of Technology, funded by the National Aeronautics and Space Administration and the National Science Foundation. It also makes use of the NASA/IPAC Infrared Science Archive and of the NASA/IPAC Extragalactic Database (NED), which are operated by the Jet Propulsion Laboratory, California Institute of Technology, under contract with the National Aeronautics and Space Administration.
LB acknowledges partial financial support from the Spanish Ministerio de Ciencia e Innovaci\'on project AYA2010-21766-C03-01.
We thank the staff at the Australia Telescope Compact Array site, Narrabri (NSW), for the valuable support. The Australia Telescope Compact Array is part of the Australia Telescope which is funded by the Commonwealth of Australia for operation as a National Facility managed by CSIRO.

\bsp

\label{lastpage}

\end{document}